\documentclass[12pt,aps,tightenlines,notitlepage,superscriptaddress]{revtex4-1}

\usepackage{amsmath}
\usepackage{amssymb}
\usepackage{bm}
\usepackage[mathscr]{eucal}
\usepackage{theorem}
\usepackage{graphicx}
\usepackage{psfrag}
\usepackage{subfigure}
\usepackage{color}
\usepackage{wasysym}
\include{graphix}
\usepackage{framed}
\usepackage{enumitem}%
\usepackage{lipsum}
\usepackage{float}
\usepackage{stmaryrd}
\usepackage{MnSymbol}
\definecolor{orange}{rgb}{1,0.5,0}

\newcommand{\mathd}{\mathrm{d}}
\newcommand{\vecx}{\bm{x}}
\newcommand{\vecu}{\bm{u}}

\newcommand{\mydiff}{\mathcal{D}}
\newcommand{\mycn}{\mathrm{Cn}}
\newcommand{\myre}{\mathrm{Re}}
\newcommand{\mype}{\mathrm{Pe}}
\newcommand{\mynu}{\nu}
\newcommand{\mydim}{n}

\newcommand{\crazyell}{s}

\begin{document}

\bibliographystyle{apsrev}

\title{A flow-pattern map for phase separation using the Navier--Stokes Cahn--Hilliard model}

\author{Aurore Naso}
\affiliation{Laboratoire de M\'ecanique des Fluides et d'Acoustique, CNRS, \'Ecole Centrale de Lyon, Universit\'e de Lyon, INSA de Lyon, 69134 \'Ecully Cedex, France}

\author{Lennon \'O N\'araigh}
\email{onaraigh@maths.ucd.ie}

\affiliation{School of Mathematics and Statistics, University College Dublin, Belfield, Dublin 4}

\date{\today}

\begin{abstract}
We use the Navier--Stokes--Cahn--Hilliard model equations to simulate phase separation with flow.  We study coarsening -- the growth of extended domains wherein the binary mixture phase separates into its component parts.  The coarsening is characterized by two competing effects: flow, and the Cahn--Hilliard diffusion term, which drives the phase separation.
Based on extensive two-dimensional direct numerical simulations, we construct a flow-pattern map outlining the relative strength of these effects in different parts of the parameter space.
The map reveals large regions of parameter space where a standard theory applies, and where the domains grow algebraically in time.  However, there are significant parts of the parameter space where the standard theory does not apply.
In one region, corresponding to low values of viscosity and diffusion, the coarsening is accelerated compared to the standard theory.
Previous studies involving Stokes flow report on this phenomenon; we complete the picture by demonstrating that this anomalous regime occurs not only for Stokes flow, but also, for flows dominated by inertia.
In a second region, corresponding to arbitrary viscosities and high Cahn--Hilliard diffusion, the diffusion overwhelms the hydrodynamics altogether, and the latter can effectively be ignored, in contrast to the prediction of the standard scaling theory.  Based on further high-resolution simulations in three dimensions, we find that broadly speaking, the above description holds there also,  although the formation of the anomalous domains in the low-viscosity-low-diffusion part of the parameter space is delayed in three dimensions compared to two.

\end{abstract}


\maketitle

\section{Introduction}

When a binary fluid in which both components are initially well mixed undergoes rapid cooling below a critical temperature, both phases spontaneously separate to form domains rich in the fluid's component parts.  The domains expand over time in a phenomenon known as coarsening~\cite{CH_orig}.  The length scale of a typical domain grows as $\ell\sim t^a$, where $\ell$ is the length scale, $t$ is time, and $a>0$ is a characteristic exponent. 
A complete mathematical model for this process is given by the Navier--Stokes Cahn--Hilliard (NSCH) equation set and involves not only the concentration field of the binary mixture but also, its velocity field~\cite{LowenTrus,ding2007diffuse}.  In this way, the coarsening can proceed via one of several mechanisms~\cite{Bray_advphys}, each possessing its own characteristic value of the exponent $a$.
Because these mechanisms arise from balancing various terms in the NSCH equations, they are also each characterized by their own dimensionless group, and the aim of the present work is mainly to characterize phase-separation in unforced turbulence (in both two and three dimensions) in the NSCH framework based on these dimensionless groups.
Our analysis reveals regions of the parameter space where the above standard scaling theory is inadequate for describing the coarsening phenomenon in both two and three dimensions.
We first of all place our work in the context of the existing literature before presenting our findings concerning anomalous coarsening dynamics.

The mathematical model used in this work is the NSCH equation set, solved on a multiply-periodic domain $[0,L]^\mydim$ (with $\mydim=2,3$), presented here in dimensional form as follows:
\begin{subequations}
\begin{eqnarray}
\frac{\partial\phi}{\partial t}+\vecu\cdot\nabla \phi&=&\frac{D}{\alpha}\nabla^2\mu,\label{eq:ch_basic}\\
\rho\left(\frac{\partial\vecu}{\partial t}+\vecu\cdot\nabla\vecu\right)&=&-\nabla p+\eta\nabla^2\vecu-\phi\nabla\mu,\qquad \nabla\cdot\vecu=0,\label{eq:ns_basic}
\end{eqnarray}%
where $\phi$ is the concentration field of the binary liquid ($\phi=\pm 1$ denotes saturation in one or other of the components whereas $\phi=0$ denotes the perfectly mixed state), and $\bm{u}$ and $p$ are the velocity and pressure fields respectively.
The constant density is denoted by $\rho$ and the constant dynamic viscosity by $\eta$ -- we assume that both phases have the same densities and dynamic viscosities.  Here also, $D$ is the diffusion coefficient and $\alpha$ is a further constant with dimensions of Energy/Volume.  
As such, the quantity $\mu$ is the chemical potential:
\begin{equation}
\mu=\alpha\left(\phi^3-\phi-\gamma\nabla^2\phi\right),
\end{equation}
\label{eq:nsch}%
\end{subequations}%
where $\sqrt{\gamma}$ is a parameter describing the width of a transition layer between typical domains.  A derivation of these equations based on physical reasoning is given in References~\cite{LowenTrus,ding2007diffuse}.  Equation~\eqref{eq:nsch} is solved in nondimensional terms, wherein the key dimensionless parameters are 
\begin{equation}
\mycn=\sqrt{\gamma/L^2},\qquad \mydiff=D/UL,\qquad \myre=UL\rho/\eta.
\end{equation}
Here, $U=\sqrt{\alpha/\rho}$ is the velocity scale, and $\mydiff$ is the inverse P\'eclet number.  As such, we solve a dimensionless version of Equation~\eqref{eq:nsch} on the unit multiply-periodic domain $[0,1]^\mydim$, making the formal replacements $D\rightarrow\mydiff$, $\eta\rightarrow\myre^{-1}$, $\gamma\rightarrow\mycn^2$, setting the parameters $\rho$ and $\alpha$ to unity and replacing the dimensional chemical potential $\mu$ by its nondimensional analogue, $\phi^3-\phi-\mycn^2\nabla^2\phi$.  The resulting dimensionless equations are solved with a prescribed initial condition $\phi(\bm{x},t=0)=\phi_0(\bm{x})$, and $\vecu(\bm{x},t=0)=0$.  The focus of the present work is on symmetric mixtures, whereby each component of the binary fluid is initially present in equal amounts, hence $\int\phi_0(\vecx)\,\mathd^\mydim x=0$.  In view of the incompressibility condition, and the ensuing flux-conservative nature of Equation~\eqref{eq:ch_basic}, it follows that $\int\phi(\vecx,t)\,\mathd^\mydim x=0$ for all $t>0$ also.

The dynamics of the Cahn--Hilliard equation~\eqref{eq:ch_basic} with flow can be classified as either passive or active -- the former refers to the case wherein the coupling term $-\phi\nabla\mu$ in Equation~\eqref{eq:ns_basic} drops out.  Both active and passive cases have been
studied extensively for two-dimensional systems, in particular for the cases of stirring by chaotic flow fields and stirring by driven two-dimensional turbulence~\cite{naraigh2007bubbles,Berti2005,shear_Berthier}.  In particular, in Reference~\cite{Berti2005} coarsening arrest in active (forced) turbulence is considered, and the arrest scale is related to the mean shear across a typical fluid domain.  Such a clear relationship between these quantities is observed in cases when the lengthscale of the arrested domains is less than the energy-injection scale.  For the opposite scenario (i.e. when the length scale of the arrested domains exceeds the energy-injection scale), the arrest scale has been shown by other researchers to be comparable to the Hinze length~\textit{et al.}~\cite{perlekar2017two}; in other words, the arrest scale in this regime is governed by a balance between inertia and surface tension (effectively, the backreaction or `active' term in the momentum equation).
Our recent work~\cite{naraigh2015flow} for the passive case in two and three dimensions has further highlighted the key role played by the P\'eclet number (measuring the strength advection term relative to the Cahn--Hilliard antidiffusion term) in the outcome of the phase separation: changing the P\'eclet number by orders of magnitude involves dramatic changes in the outcome of the phase separation.  For the active case the Reynolds number is a second key parameter.  

\begin{figure}[bht!]
	\centering
		\includegraphics[width=0.5\textwidth]{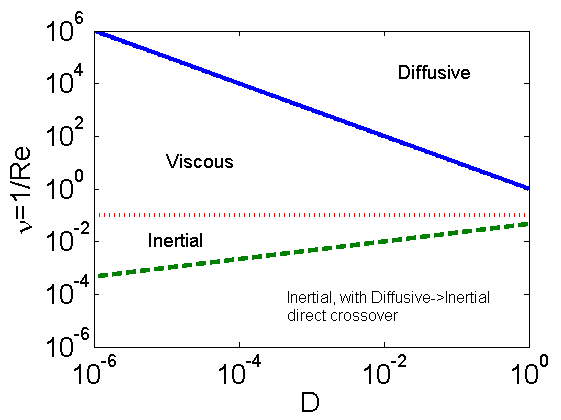}
		\caption{Notional flow-pattern map based on Equation~\eqref{eq:L_bray}.  
		The map shows the scaling regime applicable at the time where $\ell(t)=1$.  Indicative regime boundaries are shown also, corresponding to the curves 
$\nu=1/\mydiff$ (solid line) and $\nu=\sqrt{\mycn}$ (dotted line).  The lowermost curve (dashed line) corresponds to $\nu=(\mydiff\,\mycn^2)^{1/3}$.  Below this threshold, it is expected that the system will cross over from a diffusive scaling regime to an inertial one, without any intermediate viscous regime.  The value of $\mycn$ used in the plot is $10^{-2}$.}
	\label{fig:flow_pattern_map_theory}
\end{figure}

In the present work, we focus on active mixtures in both two and three dimensions.  The coupling term $-\phi\nabla\mu$ thus represents a forcing which induces hydrodynamic turbulence~\cite{Berti2005,Kendon2001}.  For this scenario, there is a simple theory based essentially on dimensional analysis that explains the scaling behaviour of the resulting domains.  The spatial dimension effectively `cancels out' in the theory; as such, the theory predicts the following three ($\mydim$-independent) distinct regimes:
\begin{itemize}
\item Diffusive scaling, with $\ell \sim t^{1/3}$.  Here, the domain formation is driven entirely by the term $\mydiff\nabla^2\mu$ in Equation~\eqref{eq:ch_basic}, and the exponent can be predicted theoretically using classical arguments~\cite{LS};
\item Viscous scaling, with $\ell \sim t$.  This regime is obtained by a balance between viscous and Korteweg stress term $-\phi\nabla\mu$ in Equation~\eqref{eq:ns_basic};
\item Inertial scaling, with $\ell \sim t^{2/3}$,  obtained by a balance between inertia and the force induced by the Korteweg stress in  Equation~\eqref{eq:ns_basic}.
\end{itemize}
By further dimensional analysis, Bray~\cite{Bray_advphys}  has identified the following conditions for the applicability of the different regimes, given here in terms of our dimensionless groups:
\begin{equation}
\ell(t)\sim\begin{cases}(\mydiff\,\mycn\, t)^{1/3},\qquad  \ell\ll \left(\mydiff/\myre\right)^{1/2},\\
\myre\,\mycn\, t,\qquad (\mydiff/\myre)^{1/2}\ll \ell \ll 1/(\myre^2\mycn),\\
(\mycn\,t^2)^{1/3},\qquad \ell\gg 1/(\myre^2\mycn).
\end{cases}
\label{eq:L_bray}
\end{equation}
As such, $\ell(t)$ is a monotone-increasing function of time, meaning that the diffusive regime should hold for very early times, the viscous regime for intermediate times, and the inertial regime for late times.  Notice that the viscous scaling regime $\ell(t)\sim \myre\,\mycn\,t$ exists only when $\mydiff\,\myre^3\,\mycn^2\ll 1$ -- otherwise the scaling behaviour crosses over from diffusive at earlier times directly to inertial at later times.  These results are summarized in Figure~\ref{fig:flow_pattern_map_theory}.

%

The above is consistent with the direct numerical simulations of Kendon et al.~\cite{Kendon2001}, who observe a crossover from viscous to inertial scaling in their 3D direct numerical simulation of the (unforced) NSCH equations.  In their simulations, a turbulent flow is observed, even for situations wherein the initial velocity is zero and the momentum has no external forcing term.  The reason for this is because the momentum equation is effectively forced by the phase separation and the `backreaction' that couples the momentum equation to the concentration gradient of the binary fluid.
The present study reveals parameter regimes where this standard scaling theory applies, as well as significant parameter regimes where it fails.  The anomalous regions exhibit interesting phenomena, not least a dramatic morphology difference compared to that usually seen in phase-separating  binary liquids.

\section{Methodology}

Equation~\eqref{eq:nsch} is solved numerically using a finite-difference code.  The spatial derivatives are treated using a MAC grid~\cite{ding2007diffuse}.  The scalar $\phi$ is defined at cell centres and the velocity components are defined at cell faces, and the convective derivatives are treated using a fifth-order WENO scheme.

In the Cahn--Hilliard equation~\eqref{eq:ch_basic}, the time-stepping is a combination of two treatments: third-order Adams--Bashforth for the convective derivative and the phase-separation term $\mydiff\nabla^2(\phi^3-\phi)$ and fully-implicit temporal discretization of the hyperdiffusion term $-\mycn^2 \mydiff\nabla^4 \phi$.  The result is a scheme that is numerically stable for reasonably large timesteps~\cite{Zhu_numerics}.  Due to the  fully-implicit temporal discretization of the hyperdiffusion term, a bilaplacian  operator must be inverted at each timestep.  This is done using the Jacobi method.  Because of the strong diagonal dominance of this operator, the method converges rapidly.  As part of an earlier work~\cite{naraigh2015flow}, the resulting Cahn--Hilliard solver has been fully validated by reference to standard benchmark tests in the literature.

The Navier--Stokes equations~\eqref{eq:ns_basic} are solved by a standard projection method \cite{Chorin68}.  The time-stepping is again a combination of different treatments: third-order Adams--Bashforth for the convective derivative and the non-diagonal part of the viscous term, and fully-implicit temporal discretization of the diagonal part of the viscous term and of the term depending on $\phi$.  The inversion of the resulting operator is done at each timestep using a mixed Jacobi/Gauss--Seidel algorithm.  More details on the method, as well as the results of some benchmark tests involving the full solver in a different context can be found in \cite{Loisy17_JFM}.

The numerical scheme has been implemented in C++ with MPI domain decomposition in each of the periodic directions.  
 All simulations carried out in this work use $\mycn^2=10^{-4}$ and a computational grid with $256^\mydim$ gridpoints.  
Grid refinement studies have been carried out in two dimensions on grids with up to $512^2$ gridpoints, by which it was confirmed that the lower grid resolution is sufficient for the present purposes.  
The timestep has been chosen in all cases so as to guarantee the stability of the code and also, the convergence of the numerical results.  
Finally, the initial condition $\phi_0(\vecx)$ is the same in all simulations, such that $\phi_0$ is chosen to be a different random number at each point $\vecx$ in the domain.  The random numbers are drawn from the uniform distribution on the interval $[-0.1,0.1]$, such that $\int_{[0,L]^\mydim} \phi_0(\vecx)\,\mathd^\mydim x=0$.

\section{Results} 
In this section we present a comprehensive set of simulation results for the two-dimensional case, culminating in a `flow-pattern map' outlining the observed domain morphology (and hence, flow regime) as a function of the dimensionless diffusivity $\mydiff$ and the effective dimensionless viscosity $\nu\equiv \myre^{-1}$.
  The analogous three-dimensional simulations are computationally more expensive; the flow-pattern map gives \textit{a priori} information to determine a subset of parameter cases to study in three dimensions.  As such, subsections~\ref{sec:prelim}--\ref{sec:anomtheor} below describe work in two dimensions, while the three-dimensional case is treated in subsection~\ref{sec:threed}.

\subsection{Preliminary results}
\label{sec:prelim}
\begin{figure}[tbh!]
\centering
\subfigure[]{\includegraphics[width=0.32\textwidth]{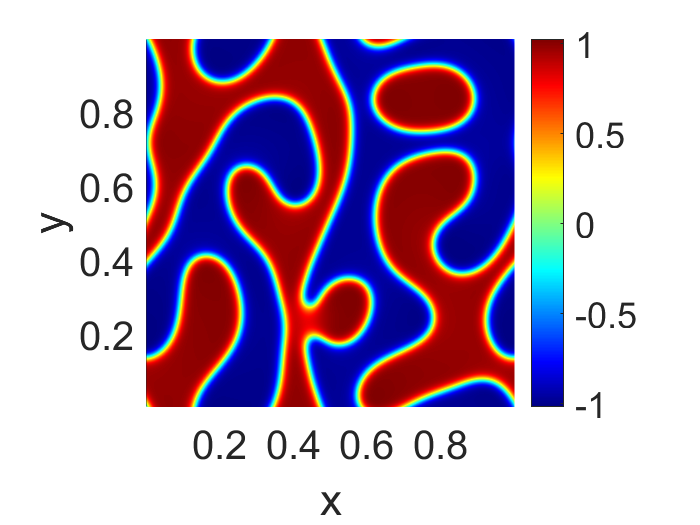}}
\subfigure[]{\includegraphics[width=0.32\textwidth]{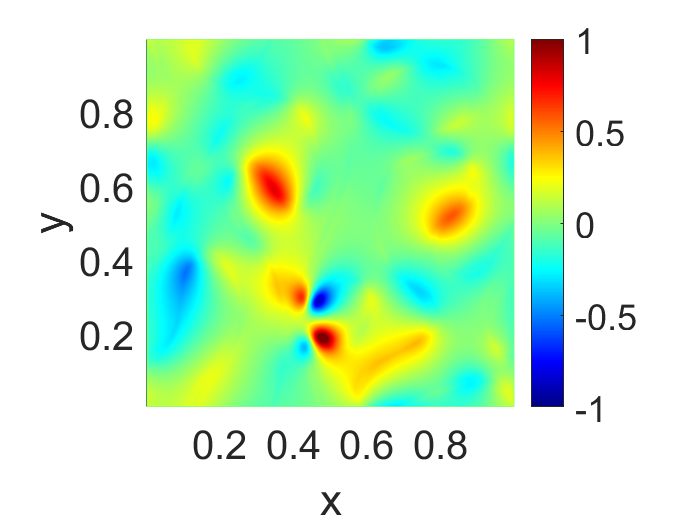}}
\subfigure[]{\includegraphics[width=0.32\textwidth]{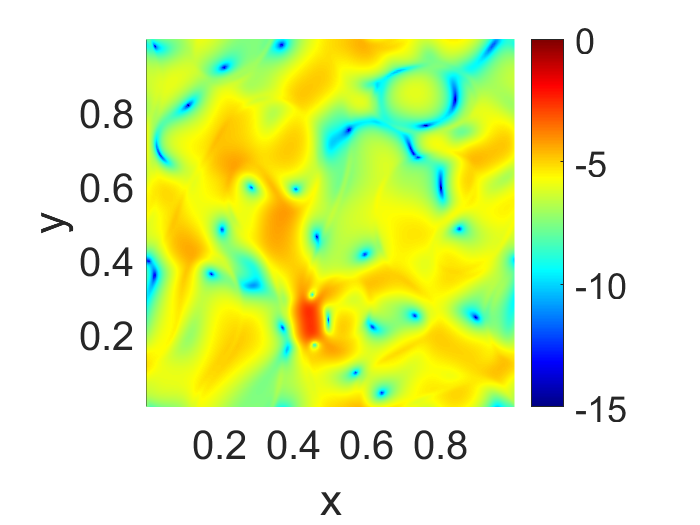}}
\caption{Case $(\mydiff,\nu)=(10^{-3},10^{-1})$: snapshots of (a) $\phi$, (b) the vorticity $\omega=\partial_x u_2-\partial_y u_1$ and (c) $\log(\varepsilon)$ (natural logarithm of the energy dissipation rate), at $t=15$.  The two-dimensional velocity field $\vecu$ is given by $\vecu=u_1\bm{e}_x+u_2\bm{e}_y$, where $\bm{e}_x$ and $\bm{e}_y$ are unit vectors in the $x$- and $y$-directions.
}
\label{fig:snapshots_run2d_15}
\end{figure}

%
%
%
%
%
%
%
We first of all present briefly some sample results for the case $\mydiff=10^{-3}$ and $\nu=10^{-1}$.  This is a preliminary discussion, the reason for which is that this parameter case illuminates the standard scaling theory and provides base cases against which to compare the other work
(these results also provide further validation of our numerical methods).
As such, typical snapshots of the distributions of $\phi$ and $\omega$ (the non-vanishing component of vorticity) are shown in Figure~\ref{fig:snapshots_run2d_15}(a-b), at $t=15$, where the inertial regime holds.  A further snapshot of $\log(\varepsilon)$ is shown at the same time in Figure~\ref{fig:snapshots_run2d_15}(c).  Here,  $\varepsilon=\myre^{-1}S_{ij}S_{ij}$ is the energy dissipation rate, with $S_{ij}=(\partial_i u_j+\partial_j u_i)/2$.
The coexistence of domains of pure phases is illustrated in panel~(a). The flow fields (panels (b) and (c)) are weakly correlated to the concentration field.

\begin{figure}[htb]
\centering
\subfigure[]{\includegraphics[width=0.45\textwidth]{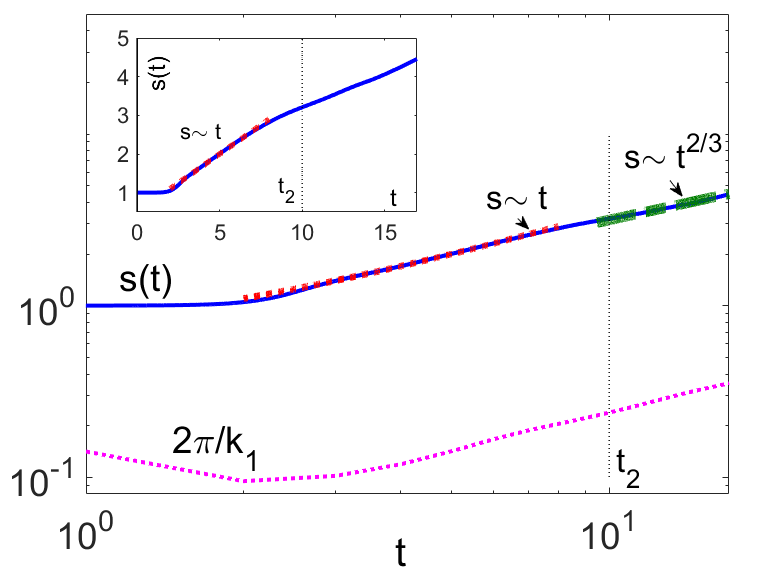}}
\subfigure[]{\includegraphics[width=0.45\textwidth]{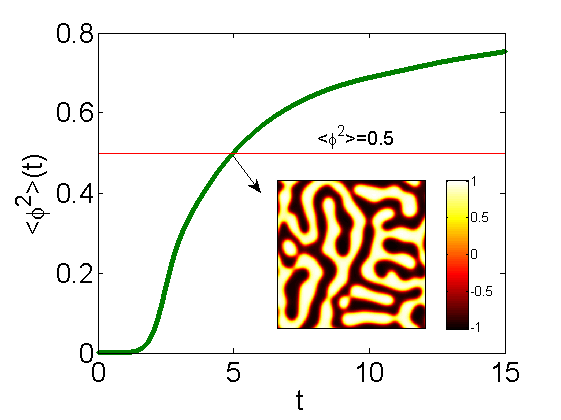}}
\caption{(a) Plot of $\crazyell(t)=(1-\langle \phi^2\rangle)^{-1}$ for the case $(\mydiff,\nu)=(10^{-3},10^{-1})$, on a log-log scale.
A further  independent measure of the domain scale, $2\pi/k_1$ is also plotted.  Shown also is the time $t_2$ where a crossover between scaling regimes is predicted, based on theory. 
The earlier crossover time $t_1$ is not shown because it is virtually indistinguishable from $t=0$.
The inset shows $s(t)$ again, plotted on a linear scale.  
(b) Plot of $\langle \phi^2\rangle$ demonstrating that the viscous scaling regime corresponds to a time where $\langle \phi^2\rangle=0.5$.  The inset shows the concentration $\phi$ at the time when $\langle \phi^2\rangle=0.5$.
}
\label{fig:lt_run2d_15_fit}
\end{figure}

\begin{figure}[htb]
\subfigure[]{\includegraphics[width=0.45\textwidth]{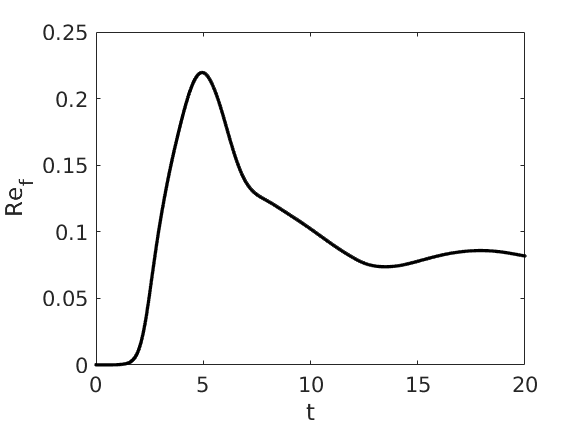}}
\subfigure[]{\includegraphics[width=0.45\textwidth]{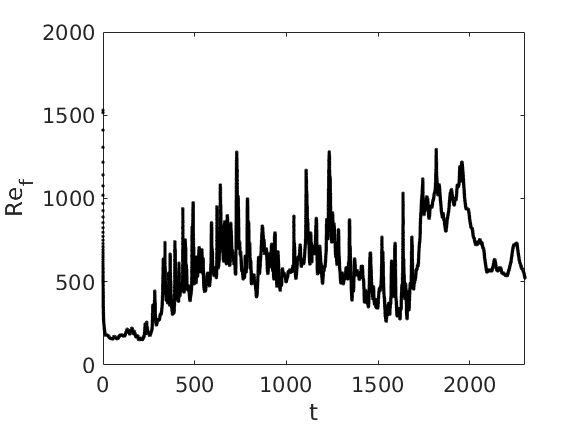}}
\caption{Time evolution of the instantaneous Reynolds number $\myre_f$ for the cases: (a) $(\mydiff,\nu)=(10^{-3},10^{-1})$; (b) $(\mydiff,\nu)=(10^{-5},10^{-5})$.}
\label{fig:ref}
\end{figure}

To quantify the coarsening rate of the domains seen in Figure~\ref{fig:snapshots_run2d_15}, the quantity $\crazyell(t):=(1-\langle \phi^2\rangle)^{-1}$ is plotted as a function of time in Figure~\ref{fig:lt_run2d_15_fit}(a), on a log-log scale.  Here, the angle brackets denote spatial averaging.  This quantity is a suitable proxy measure of the length scale $\ell(t)$ of a typical domain (see Reference~\cite{Kendon2001} and also, \ref{sec:app:proxy}) -- this is verified in Figure~\ref{fig:lt_run2d_15_fit}(a) also, where a comparison between $\crazyell(t)$ and a further independent measure of the domain scale, $2\pi/k_1$ is shown, with good agreement (up to a prefactor) between the two.  Here,
\begin{equation}
k_1=\frac{\int \mathd^\mydim k |\widehat\phi_{\bm{k}}|^2}{\int \mathd^\mydim k |\bm{k}|^{-1} |\widehat\phi_{\bm{k}}|^2},
\label{eq:k1def}
\end{equation}
where $\widehat{\phi_{\bm{k}}}$ is the Fourier transform of $\phi(\vecx,t)$, with $\widehat{\phi_{\bm{0}}}=0$ for the present symmetric mixture.
The results show very clearly a crossover between a regime of viscous scaling and a regime of inertial scaling.  To test the theory further, we have computed the time of the expected crossovers from Equation~\eqref{eq:L_bray}: a diffusive-viscous crossover is expected at $t=t_1$, where $(\mydiff\,\mycn\,t_1)^{1/3}=\mycn\,\myre\,t_1$, while a viscous-inertial crossover is expected at $t=t_2$, where $\mycn\,\myre\,t_2=(\mycn\,t_2^2)^{1/3}$.  Based on the theory, the diffusive-viscous crossover takes place at a very early time ($t_1=0.1$): the system is practically in a viscous scaling regime from the beginning, consistent with  Figure~\ref{fig:lt_run2d_15_fit}(a).  The very clear viscous-inertial crossover in Figure~\ref{fig:lt_run2d_15_fit}(a), at $t_2=10$, is further consistent with the theory.  
The crossover into an inertia-dominated regime is consistent also with an inertial regime, as understood in the theory of turbulence: in this regime ($t\gtrsim t_2=10$), the flow is found to be statistically isotropic, and stationary -- as evidenced in Figure~\ref{fig:ref}(a), wherein 
the Reynolds number $\myre_f(t)\equiv u_{\mathrm{rms}}(t)L\rho/\eta$  is shown to reach a steady state (here, $u_{\mathrm{rms}}$ is the root mean square of a velocity component).


\subsection{Flow-pattern map}

We carry out a range of simulations for different values of $\mydiff$ and $\myre$ and plot the results in a flow-pattern map, with a view to comparing the results with the standard scaling theory in Equation~\eqref{eq:L_bray}.
We determine from the numerical results the scaling regime exhibited by the system when the binary-fluid domains are comparable to the box size (before finite-size effects occur).
A robust quantitative criterion for this event is found by inspection to be $\langle \phi^2\rangle=1/2$.  Although theory tells us that $\crazyell(t)=(1-\langle \phi^2\rangle)^{-1}$ is a suitable proxy for the domain scale only when $\phi =\pm 1$ (and hence, when $\langle \phi^2\rangle\approx 1$), we verify \textit{a posteriori} that this criterion is appropriate more generally; e.g. in Figure~\ref{fig:lt_run2d_15_fit} the system already possesses a clear domain structure and $s(t)$ visibly demonstrates the viscous scaling behaviour when $\langle \phi^2\rangle\approx 0.5$.

 As such, a time interval corresponding to 
$\langle \phi^2\rangle\approx 1/2$ is examined for each of the parameter cases and the corresponding scaling behaviour $\crazyell(t) \sim t^{a}$ is identified, with $a=1/3$ for diffusive, $a=1$ for viscous, and $a=2/3$ for inertial regimes.
 This approach makes sense physically, as we thereby identify the scaling regime that applies just before the onset of finite-size effects, which are not taken into account in the standard scaling theory.  At the same time, it is sensible to recognize the box size as an important parameter (implicitly appearing in our simulations via the dimensionless groups), as any real system is finite in extent.  
\begin{figure}[htb]
	\centering
		\includegraphics[width=0.85\textwidth]{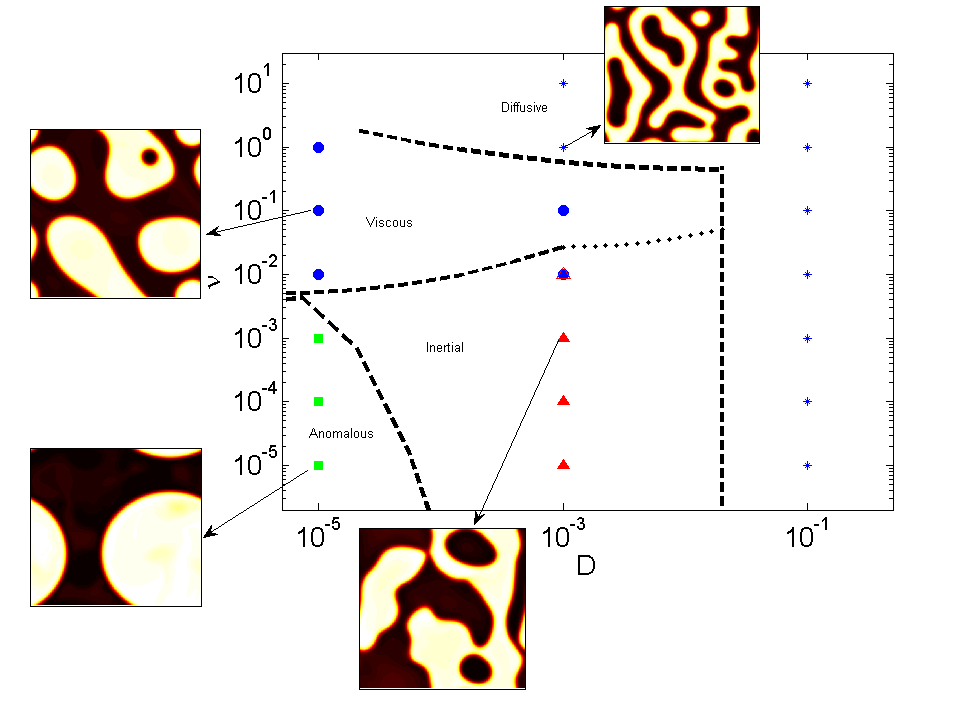}
		\caption{Flow-pattern map for the simulations when $\langle\phi^2\rangle=1/2$.  The stars in the figure denote diffusive cases, the filled circles denote viscous cases, and the filled triangles denote inertial cases.  A fourth (anomalous) case is identified also, and is discussed in the text.  The point marked by both filled circle and filled triangle corresponds to a viscous/inertial crossover at $\langle\phi^2\rangle=1/2$.		
The thick dashed lines denote indicative boundaries of the different regimes in the flow-pattern map; the dotted line segment indicates an ambiguous boundary due to the 
viscous/inertial crossover at $\langle\phi^2\rangle=1/2$.
Snapshots of $\phi$ illustrate the concentration field topology in the four regimes.}
	\label{fig:flowmap}
\end{figure}

The results of this exercise are shown in Figure~\ref{fig:flowmap}.  Four flow regimes are identified, each characterized by a distinct growth law for $\crazyell(t)$ and also, by the morphology of the domains.  For large values of the viscosity, $\crazyell(t)\sim t^{1/3}$ corresponding to a diffusive regime. The domain morphology is the classical interconnected structure observed for diffusive scaling of symmetric binary mixtures.   For small values of the viscosity (and for intermediate values of $\mydiff$), $\crazyell(t)\sim t^{2/3}$, corresponding to an inertial regime.  The domain  morphology is again an interconnected structure, albeit that the interfaces are more irregular, consistent with the presence of turbulence in the corresponding velocity field.  This is also consistent with earlier numerical simulations of the same system~\cite{wagner1998breakdown}.  Finally, at intermediate values of the dimensionless viscosity $\nu$ and small or intermediate values of the diffusion parameter $\mydiff$, there is a viscous regime, wherein $\crazyell \sim t$.  Here, the domain morphology is more droplet-like, consistent with a strong hydrodynamic effect, wherein the backreaction term in the NSCH equations effectively acts as a surface tension, promoting more spherical-like droplets~\cite{vladimirova1999two}.  

These aspects of the flow-pattern map are consistent with the classical scaling theory as summarized graphically in Figure~\ref{fig:flow_pattern_map_theory}.  However, two large areas of the flow-pattern map are not consistent with the theory.  The first is the large-$\mydiff$ region of the map: the numerical simulations show this to be unconditionally diffusive: there is no diffusive-inertial crossover at sufficiently large values of $\mydiff$.  The reason is that the diffusive behaviour in the Cahn--Hilliard equation simply overwhelms the flow, such that the morphology is determined entirely by the former.  The second is the small-$\mydiff$-small-$\mynu$ region of the parameter space, where the morphology consists of entirely spherical droplets, and where $\crazyell(t)$ does not grow as a power of $t$.  For the present purposes, we call this an anomalous regime, which we investigate further below.

\subsection{Anomalous regime -- qualitative description}
\label{sec:anom}
\begin{figure}[htb]
	\centering
		\subfigure[$\,\,t=10$]{\includegraphics[width=0.3\textwidth]{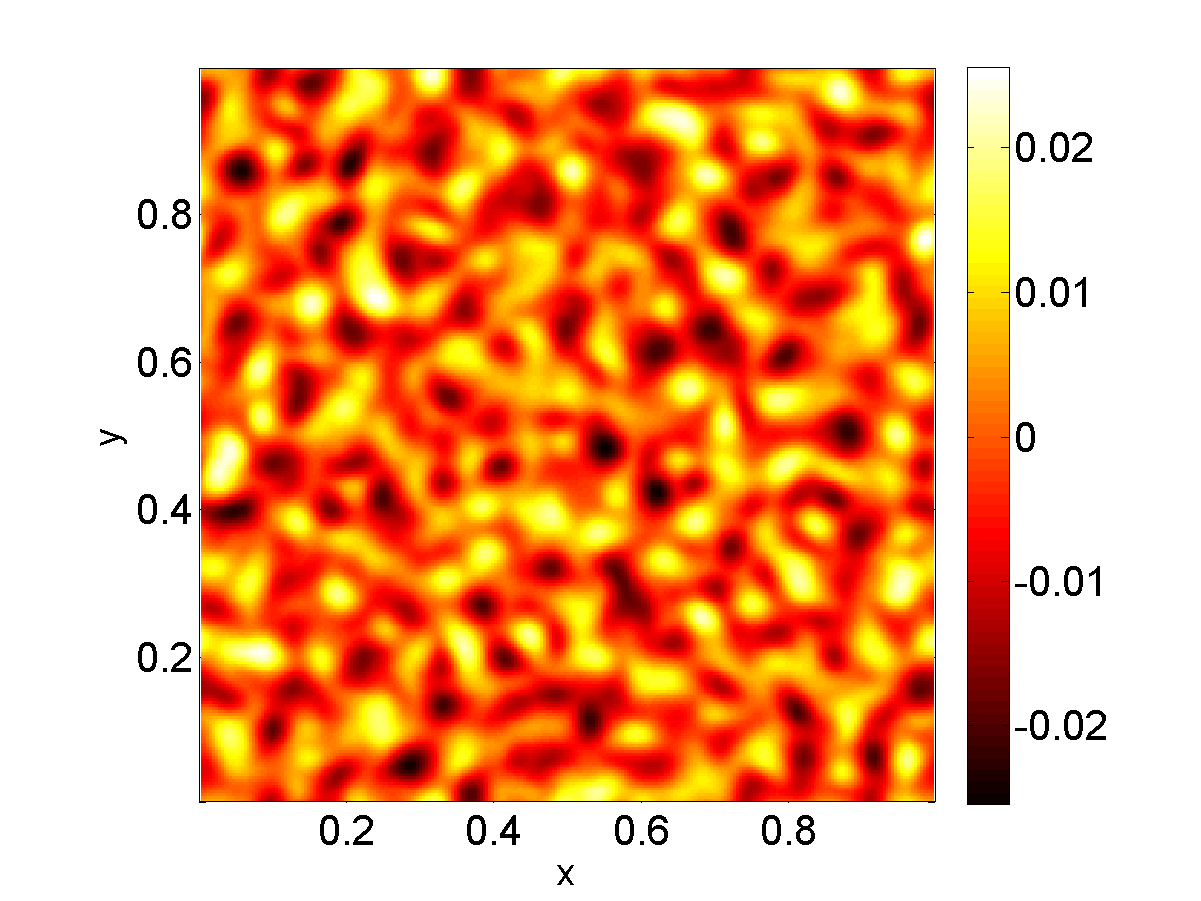}}
		\subfigure[$\,\,t=100$]{\includegraphics[width=0.3\textwidth]{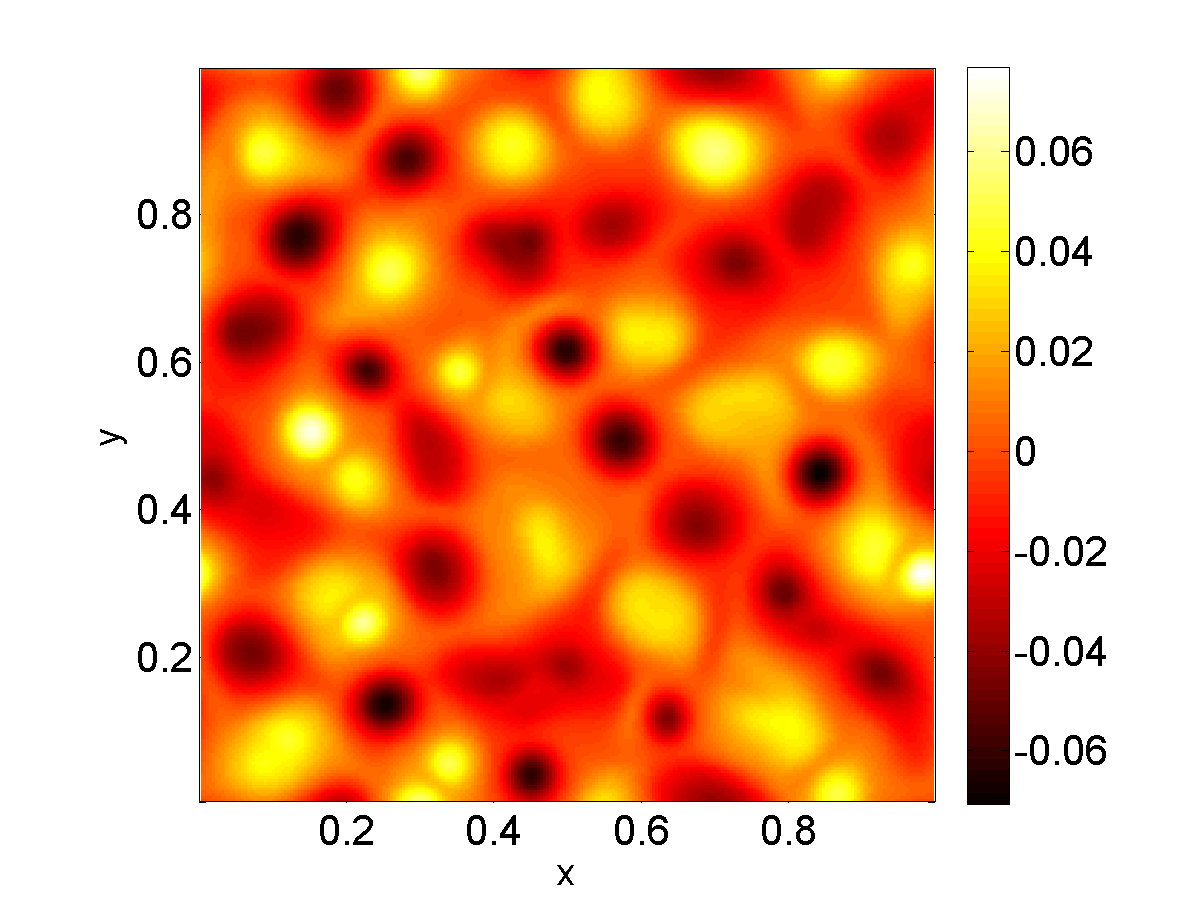}}
		\subfigure[$\,\,t=200$]{\includegraphics[width=0.3\textwidth]{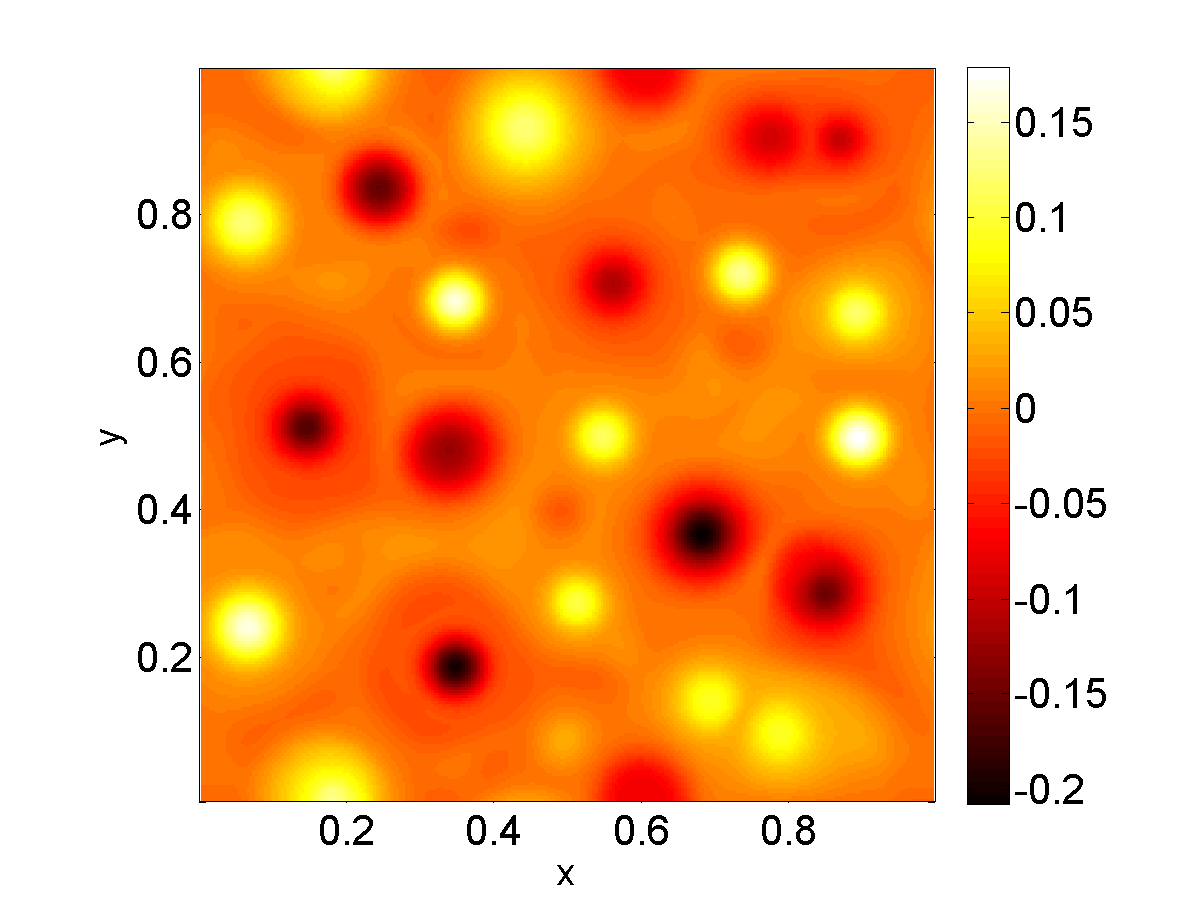}}
\caption{Snapshots of $\phi$ at early times for the case $\mydiff=\nu=10^{-5}$.}
\label{fig:phi_anomalous1}
\end{figure}

\begin{figure}[htb!]
	\centering
		\subfigure[$\langle\phi^2\rangle=0.124, t=900$]{\includegraphics[width=0.4\textwidth]{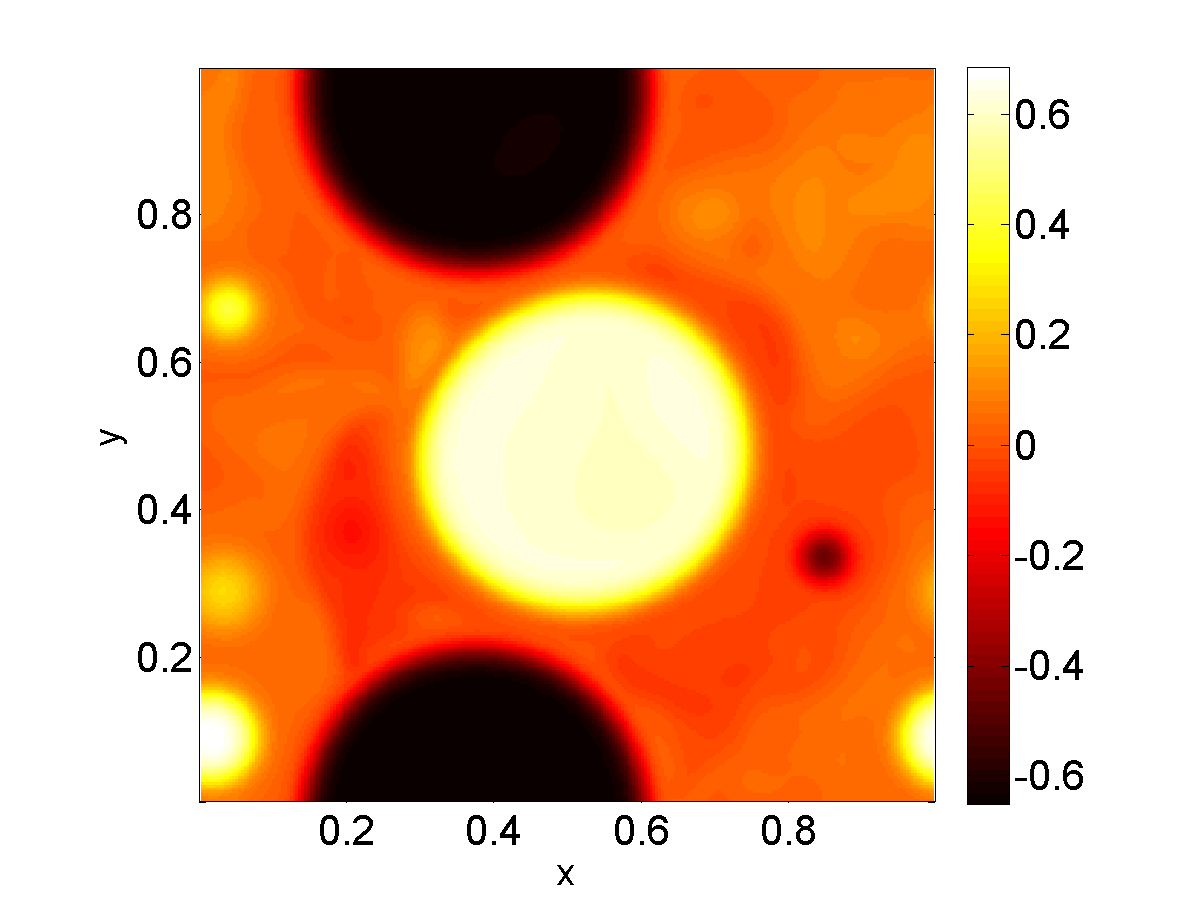}}
		\subfigure[$\langle\phi^2\rangle=0.124, t=212.5$]{\includegraphics[width=0.4\textwidth]{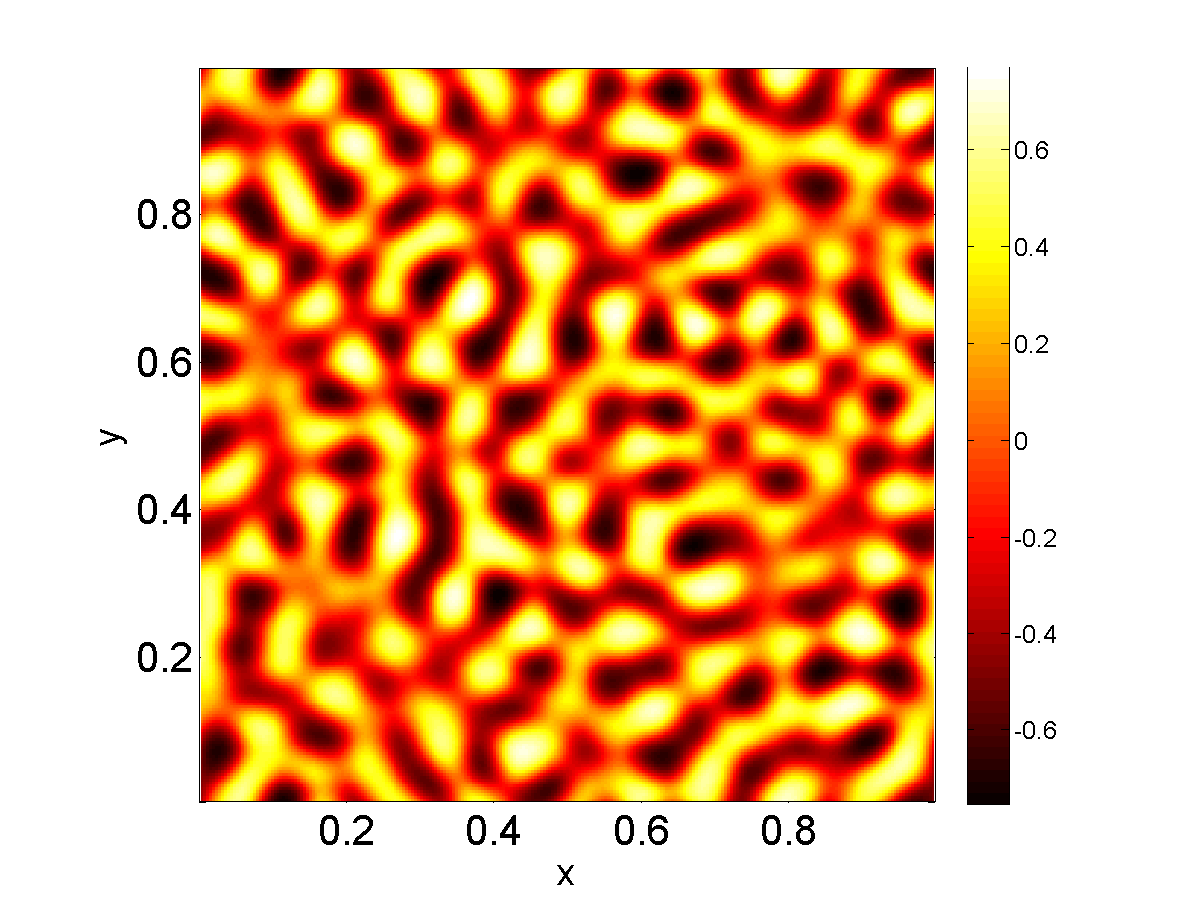}}\\
		\subfigure[$\langle\phi^2\rangle=0.467, t=1800$]{\includegraphics[width=0.4\textwidth]{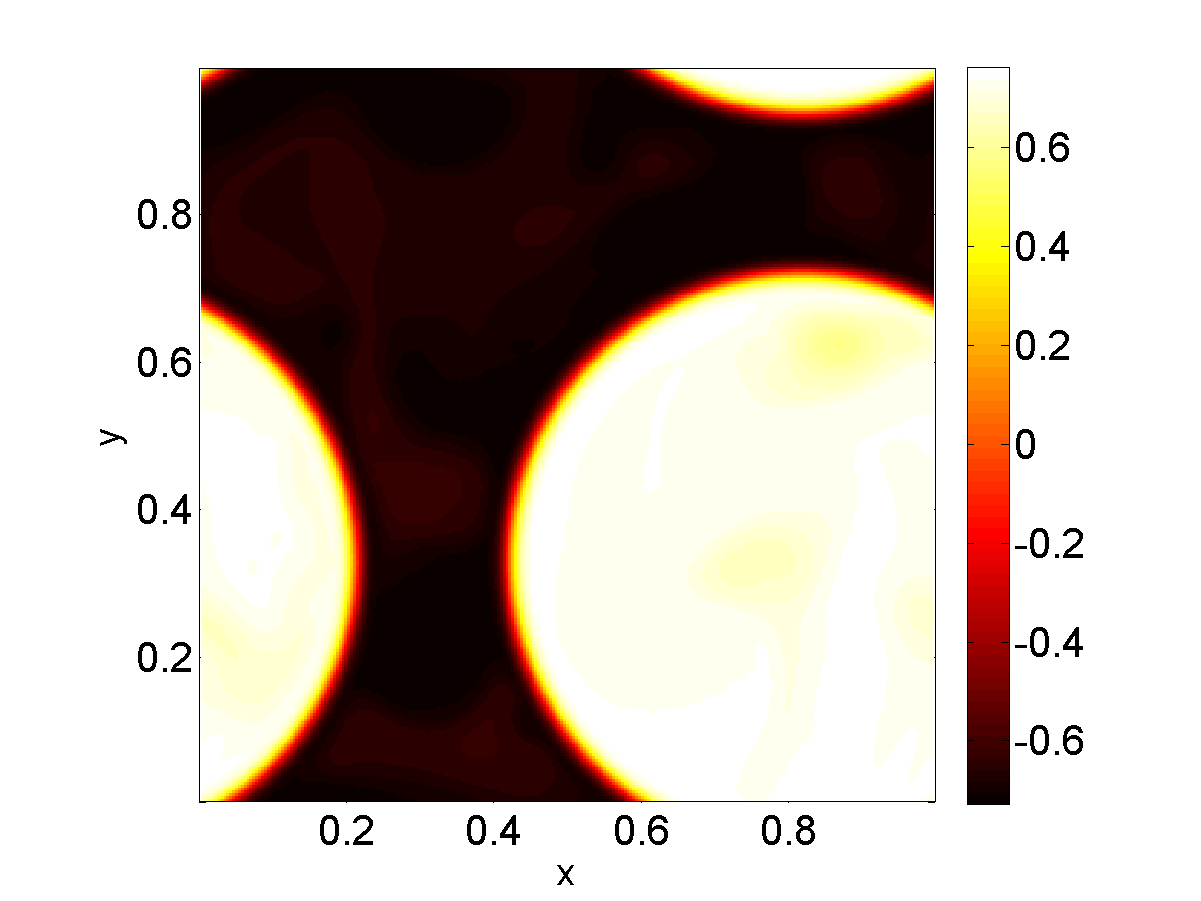}}
		\subfigure[$\langle\phi^2\rangle=0.467, t=320$]{\includegraphics[width=0.4\textwidth]{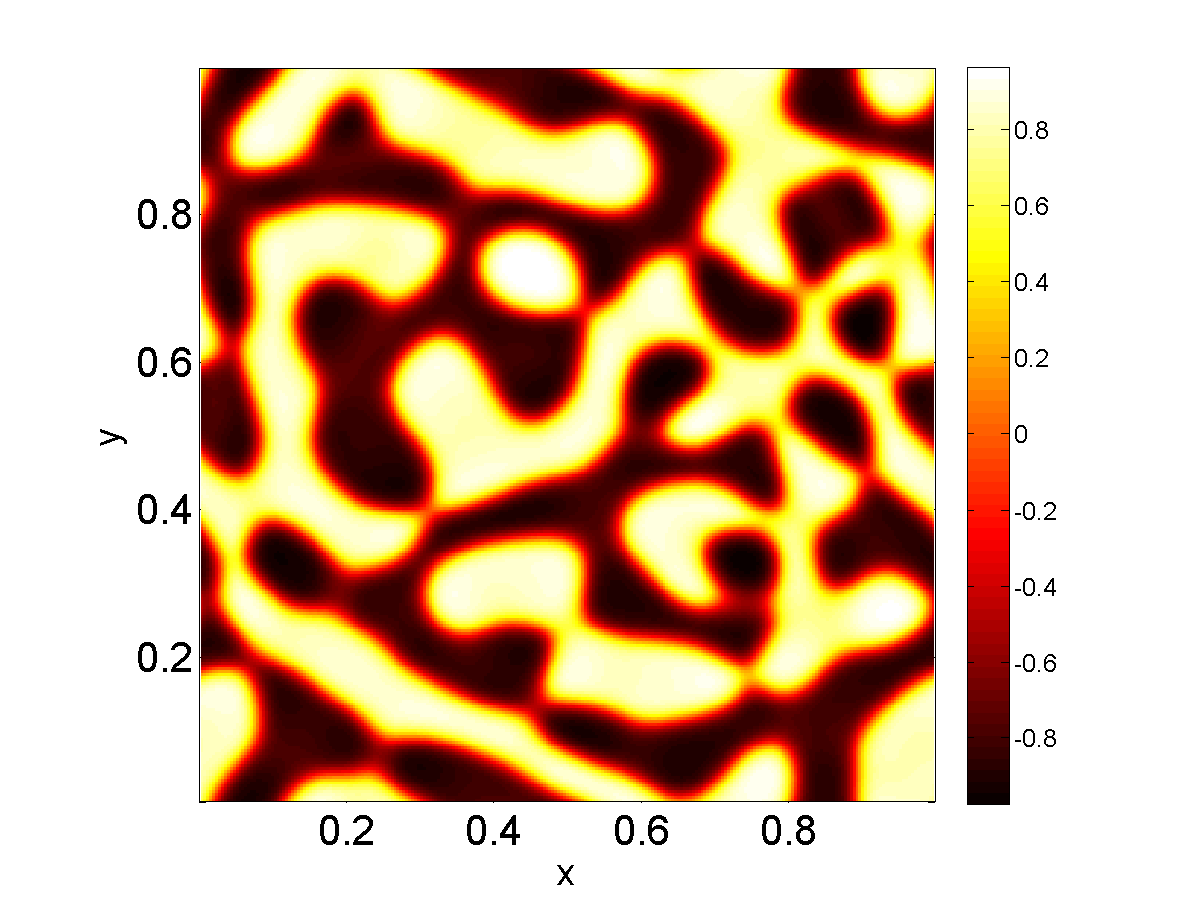}}\\
		\subfigure[$\langle\phi^2\rangle=0.758, t=2330$]{\includegraphics[width=0.4\textwidth]{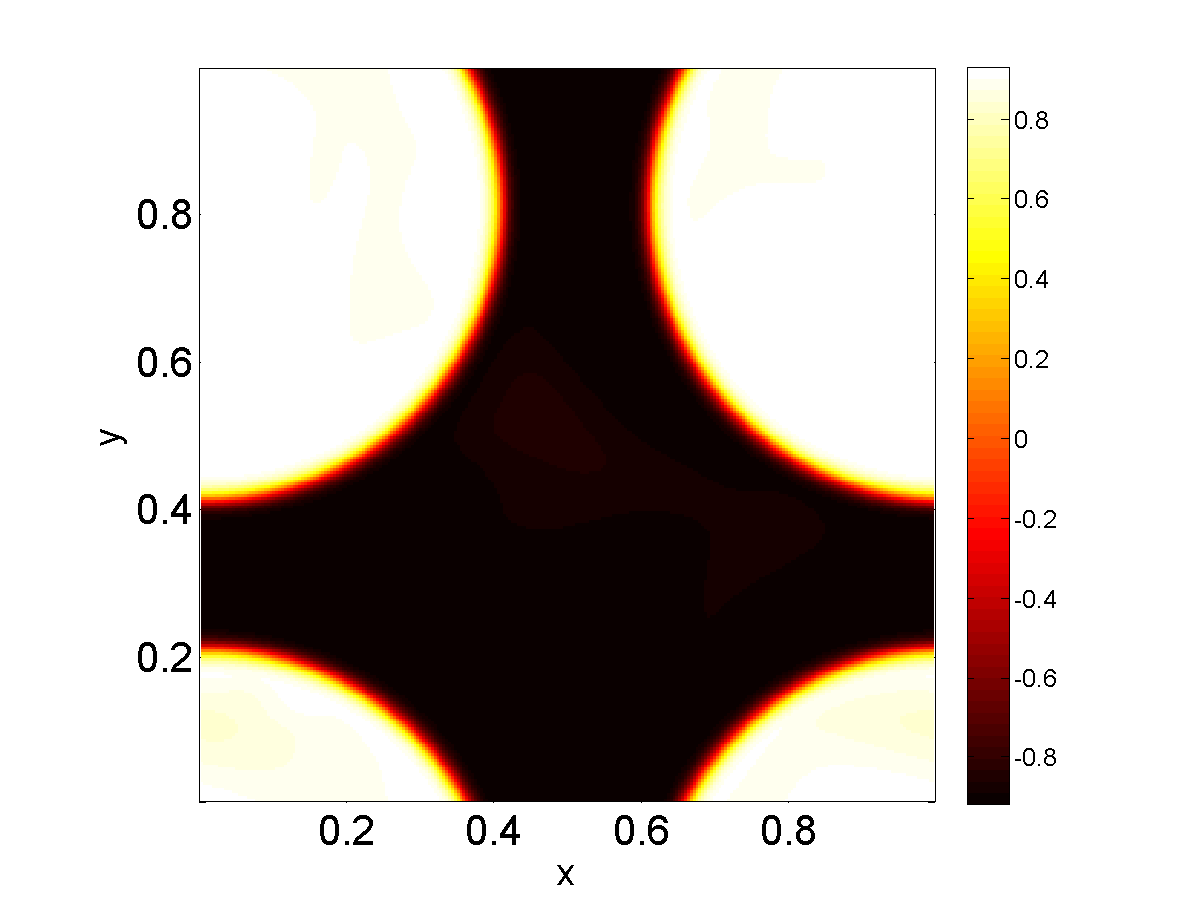}}
		\subfigure[$\langle\phi^2\rangle=0.758, t=582.5$]{\includegraphics[width=0.4\textwidth]{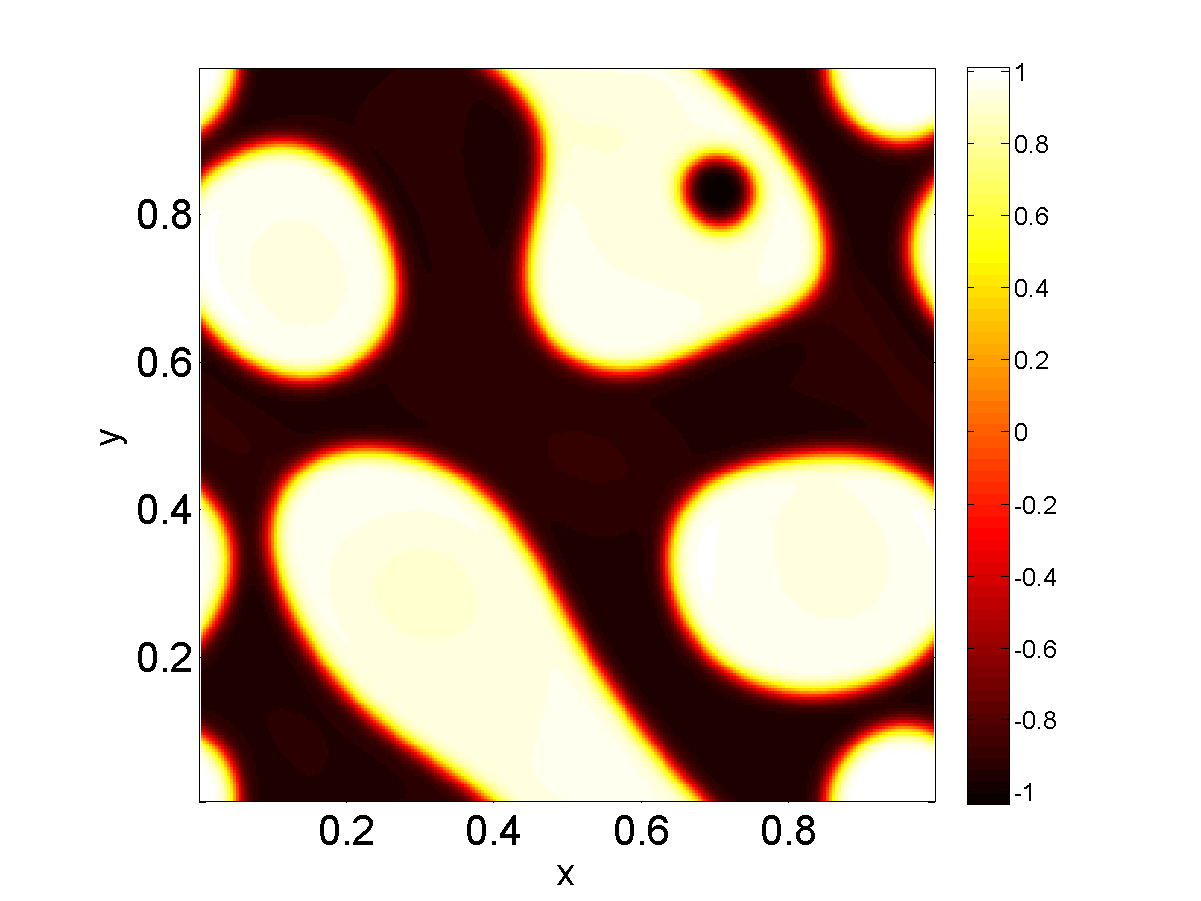}}
\caption{Snapshots of $\phi$.  First column: the case $\mydiff=\nu=10^{-5}$; second column: $(\mydiff,\nu)=(10^{-3},1)$.  The snapshots are made on the basis of equal values of $\langle\phi^2\rangle$.  Thus, in panel (a), $t=900$, in panel (b) $t=212.5$, such that $\langle \phi^2\rangle_{\text{panel (a)}}=\langle\phi^2\rangle_{\text{panel (b)}}$, and similarly for panels (c)--(f).}
\label{fig:phi_anomalous2}
\end{figure}

\begin{figure}[htb!]
	\centering
		\includegraphics[width=0.65\textwidth]{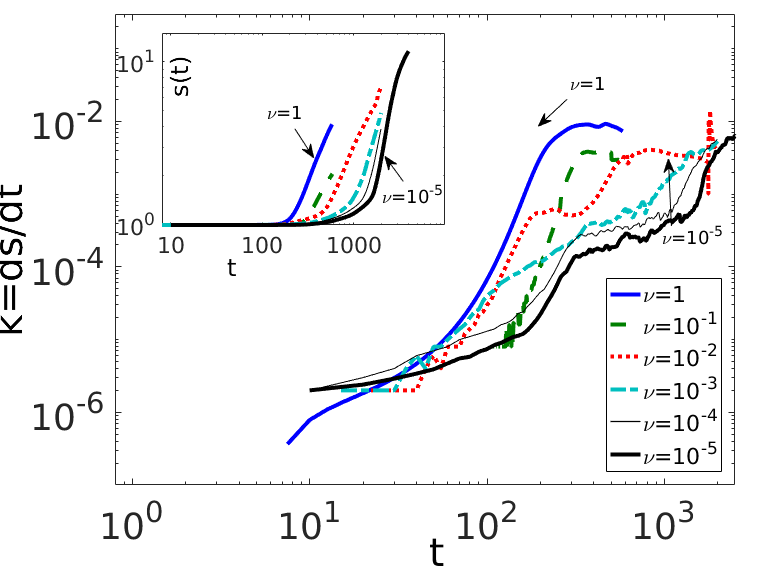}
\caption{Plot of $\mathd\crazyell/\mathd t$, where $\crazyell(t)=(1-\langle\phi^2\rangle)^{-1}$, for $\mydiff=10^{-5}$ and several values of $\nu$.  Inset: $\crazyell(t)$.  The sharp uptick in $\mathd s/\mathd t$ at $t\approx 1770$ and $\nu=10^{-2}$ is due to the sudden appearance of finite-size effects in that simulation.}
	\label{fig:lt_compare}
\end{figure}

%
%
%
The low-viscosity-high-P\'eclet-number region in Figure~\ref{fig:flowmap} exhibits this anomalous regime where the standard scaling theory is no longer in evidence.  As such, snapshots of $\phi$ for early times are shown in Figure~\ref{fig:phi_anomalous1}  for the case $\mydiff=\nu=10^{-5}$.
Further snapshots of $\phi$ at later times are shown in Figure~\ref{fig:phi_anomalous2}, where the morphology is compared with that of the standard viscous regime, with $(\mydiff,\nu)=(10^{-3},1)$.  The comparison is done on the basis of equal values of $\langle \phi^2\rangle$, as described in the caption.  
%
%
%
We also compare the quantity $\crazyell(t):=(1-\langle \phi^2\rangle)^{-1}$ for the anomalous cases with standard viscous cases, at the same value of $\mydiff$ but for a range of values of $\nu$.  The results are shown in the inset in Figure~\ref{fig:lt_compare}.  
The main part of the figure shows $\mathd \crazyell/\mathd t$ -- only for the viscous case ($\mynu\gtrsim 10^{-2}$) does this quantity saturate, indicating classical viscous scaling behaviour $\crazyell\sim t$.  For the other cases, $\crazyell(t)$ grows faster-than-linearly.

\begin{figure}[htb!]
\centering
\subfigure[]{\includegraphics[width=0.49\textwidth]{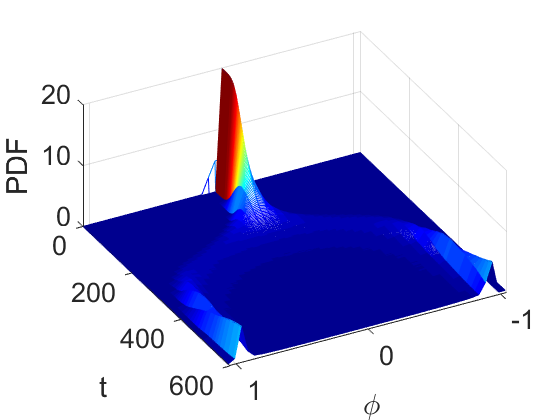}}
\subfigure[]{\includegraphics[width=0.49\textwidth]{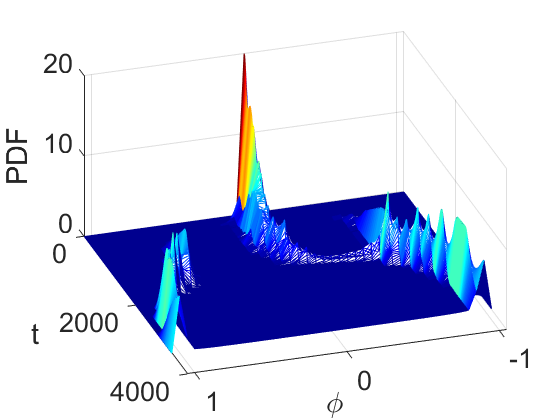}}
\caption{Time evolution of the probability distribution function of the scalar concentration $\phi$ for (a) the viscous case and (b) the anomalous case.  Parameter values:
$(\mydiff,\nu)=(10^{-3},1)$ in (a) and $\mydiff=\nu=10^{-5}$ in (b). }
\label{fig:hist_compare}
\end{figure}

We finally compare in Figure~\ref{fig:hist_compare} the time evolution of the probability distribution function (PDF) of $\phi$ of the selected viscous case $(\mydiff,\nu)=(10^{-3},1)$ with the anomalous case.  Panels (a) and (b) show the time-evolution of the PDF for the viscous and anomalous cases, respectively.
In panel (a) the process of domain formation is characterized by a rapid destabilization of the state $\phi=0$ (consistent with the linear instability of the state $\phi=0$ to small-amplitude perturbations).  Thereafter, extended regions where $\phi= \pm 1$ form, in a symmetric manner (see also Fig. \ref{fig:phi_anomalous2}(b,d,f)).  In contrast, in panel (b) the process of droplet formation is accompanied by an initial meta-stabilization of the state $\phi=0$.  Thereafter, extended regions where $\phi=\pm 1$ form, while long-lived mixed regions with $\phi=0$ persist (see also Fig. \ref{fig:phi_anomalous1}(c) and \ref{fig:phi_anomalous2}(a)) until late times whereupon the droplets fill out the entire computational domain (Fig. \ref{fig:phi_anomalous2}(c,e)).  The destabilization of the mixed state occurs via a kind of `symmetry breaking' whereby the long-lived mixed regions transform spontaneously into $\phi\approx -1$ domains -- as evidenced by the time evolution of the histogram therein.  The symmetry-breaking takes place within the overall context of mass-conservation, whereby $\langle\phi\rangle$ remains zero for all time. 

\begin{figure}[htb!]
\centering
\subfigure[]{\includegraphics[width=0.41\textwidth]{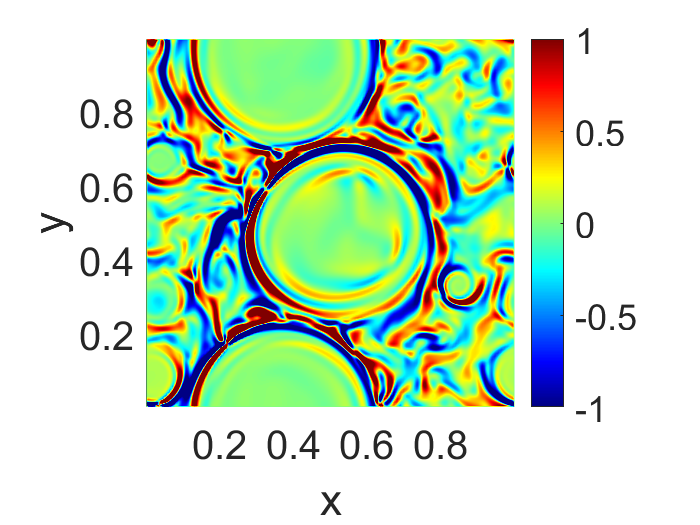}}
\subfigure[]{\includegraphics[width=0.41\textwidth]{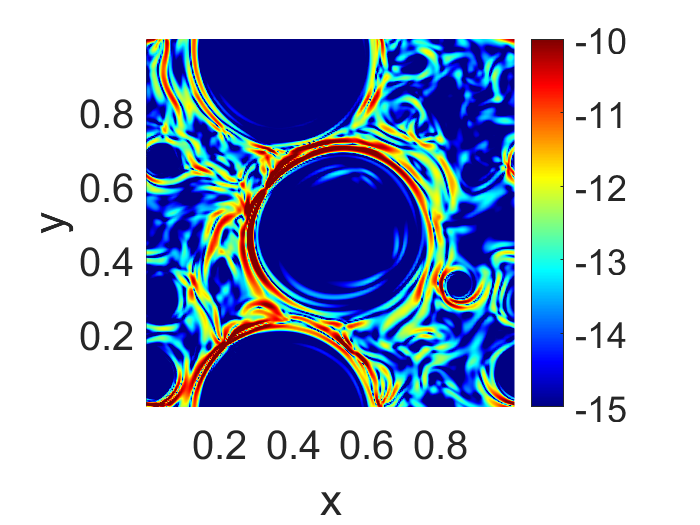}}
\subfigure[]{\includegraphics[width=0.41\textwidth]{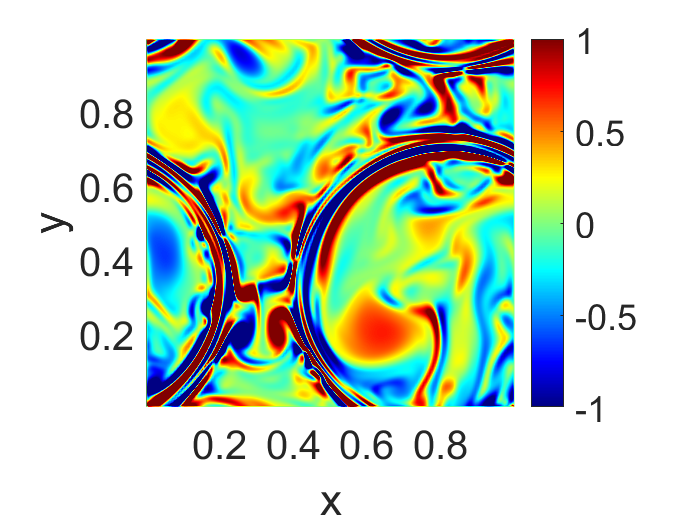}}
\subfigure[]{\includegraphics[width=0.41\textwidth]{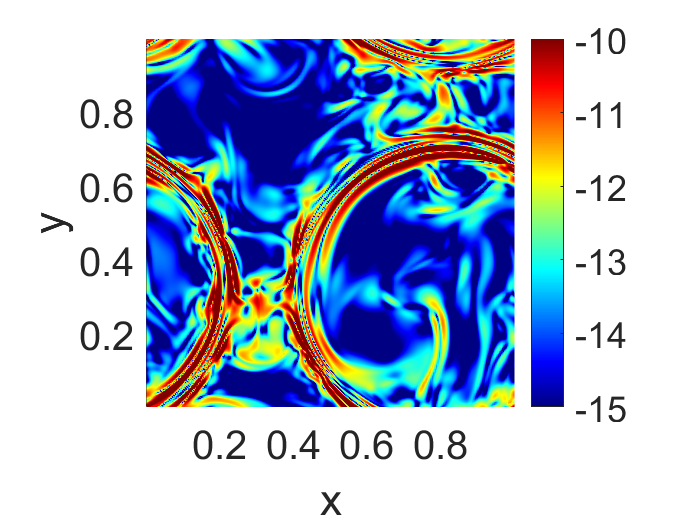}}
%
\caption{Case $(\mydiff,\myre^{-1})=(10^{-5},10^{-5})$: snapshots of (a,c) the vorticity and (b,d) $\log(\varepsilon)$ (energy dissipation rate).
Top row: $t=900$, as in Figure~\ref{fig:phi_anomalous2}(a).
Bottom row: $t=1800$, as in Figure~\ref{fig:phi_anomalous2}(c).
%
%
}
\label{fig:flow_anomalous}
\end{figure}

Next, we examine the flow in the anomalous regime.  This is illustrated in Figure~\ref{fig:flow_anomalous}, in which are plotted the vorticity and energy dissipation fields at $t=900$ and $t=1800$. Comparing them with the concentration fields at the same times (see Figures~\ref{fig:phi_anomalous2}(a) and (c)) shows that, contrary to what occurs in the inertial regime (see Figure~\ref{fig:snapshots_run2d_15}), the flow and the concentration fields are then strongly correlated. In particular, the vorticity and the energy dissipation are particularly intense in the regions of high concentration gradients, that is in the transition layers between domains of pure and mixed phases ($t=900$) or of different pure phases ($t=1800$).  In the anomalous regime, the Korteweg stress term therefore acts as a vorticity source. 
Like in the viscous regime, the flow induced by the phase separation is found to be statistically isotropic and stationary, the latter property being illustrated in Figure~\ref{fig:ref}(b).
%
%

\subsection{Anomalous regime -- theoretical understanding}
\label{sec:anomtheor}

The anomalous region of the parameter space  can be understood in a theoretical framework by breaking up the regular-to-anomalous transition into two separate routes:
\begin{itemize}
\item A viscous-to-anomalous transition, which occurs for example at $\mydiff=10^{-5}$ and as $\mynu$ is reduced from $10^{-2}$ to $10^{-3}$.
\item An inertial-to-anomalous transition, which occurs for example at $\mynu=10^{-5}$ and as $\mydiff$ is reduced from $10^{-3}$ to $10^{-5}$.
\end{itemize}
The first of these routes (viscous-to-anomalous) can be understood in the context of the existing literature phase separation driven by pure Stokes flow (i.e. no inertial term in the hydrodynamics), where a similar anomalous scaling regime has been observed previously, both numerically~\cite{vladimirova1999two,wagner1998breakdown}, and experimentally~\cite{tanaka1994double}.  The basic insight in this prior work is that the anomalous behaviour is the result of a breakdown in a previously-assumed separation of timescales. For, in the other (regular) regimes, the phase separation happens very rapidly, such that the segregated domains wherein $\phi=\pm 1$ form rapidly, and are thereafter modified by the flow on a much slower timescale.  In contrast, in the anomalous regime, the timescales for the flow and for the phase separation are comparable, such that their effects occur together~\cite{tanaka1994double}.  This explains the droplet formation in Figure~\ref{fig:phi_anomalous2}(a,c,e): as the phase separation occurs, the backreaction acts simultaneously as surface-tension-like force to promote circular droplets.  

From this point of view, it can be argued that the (viscous) hydrodynamic timescale is derived by balancing the viscous and Korteweg stress terms (as the balance between hydrodynamics and phase separation happens at early times, before the onset of inertial effects), giving a timescale $T_\mathrm{visc-hyd}=\eta/\alpha$, while the phase-separation timescale is $T_{\mathrm{ps}}=\gamma/D$.  Therefore, in order for the hydrodynamic effects to dominate the purely diffusive phase-separation effects from the beginning of the simulation, we require $T_\mathrm{visc-hyd}/T_{\mathrm{ps}}\ll 1$, hence $D\eta/\alpha\gamma \ll 1$, or $\mydiff\nu\ll \mycn^2$.   This is consistent with the flow-pattern map in Figure~\ref{fig:flowmap}, where a viscous-to-anomalous transition  occurs  at $\mydiff=10^{-5}$ and as $\mynu$ is reduced from $10^{-2}$ to $10^{-3}$.
Notice finally that the ratio $T_\mathrm{visc-hyd}/T_{\mathrm{ps}}$ is an inverse P\'eclet number based on a velocity scale $V_{\mathrm{visc}}=\alpha\sqrt{\gamma}/\eta$, hence 
\begin{equation}
\frac{T_{\mathrm{visc-hyd}}}{T_\mathrm{ps}}=\frac{D\eta}{\alpha\gamma}=\frac{D}{V_{\mathrm{visc}}\sqrt{\gamma}}:=\mype_\mathrm{visc}^{-1},
\end{equation}
and hence, equivalently, $\mype_\mathrm{visc} \gg 1$ for the onset of the anomalous regime.  This is consistent with the condition for the onset of the anomalous regime in pure Stokes flow~\cite{vladimirova1999two}.

The second of the routes can be understood similarly, by comparing the inertial timescale $T_{\mathrm{in-hyd}}=\sqrt{\rho\gamma/\alpha}$ to the phase-separation timescale $T_{\mathrm{ps}}$.  As such, for the hydrodynamic effects to dominate the purely diffusive phase-separation effects from the beginning of the simulation, we require $T_\mathrm{in-hyd}/T_{\mathrm{ps}}\ll 1$, hence 
$\sqrt{\rho/\alpha}(D/\sqrt{\gamma})\ll 1$, or $\mydiff\ll \mycn$.  This is consistent with the flow-pattern map~\ref{fig:flowmap}, where an inertial-to-anomalous transition  occurs  at fixed $\mynu=10^{-5}$ and as $\mydiff$ is reduced from $10^{-3}$ to $10^{-5}$.
We emphasize finally that these arguments advance the understanding in the current literature (e.g.  Reference~\cite{vladimirova1999two}), as we hereby demonstrate that a regular-to-anomalous transition can occur not only for Stokes flow, but also, for the full Navier--Stokes flow. In particular, we have shown that there exists a transition from an inertial scaling regime to an anomalous one.

\subsection{Extension to three dimensions}
\label{sec:threed}

\begin{figure}[htb!]
	\centering
		\subfigure[Run3D\_1]{\includegraphics[width=0.8\textwidth]{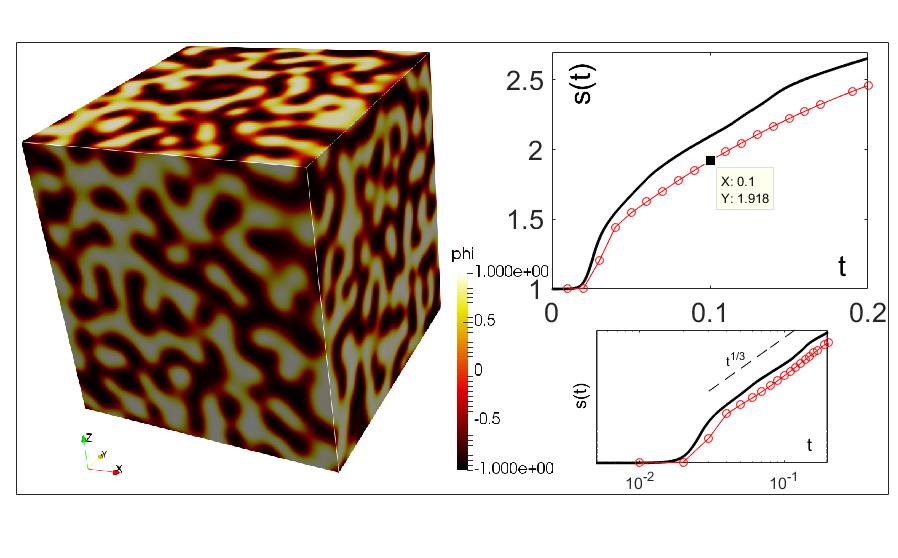}}
		\subfigure[Run3D\_2]{\includegraphics[width=0.8\textwidth]{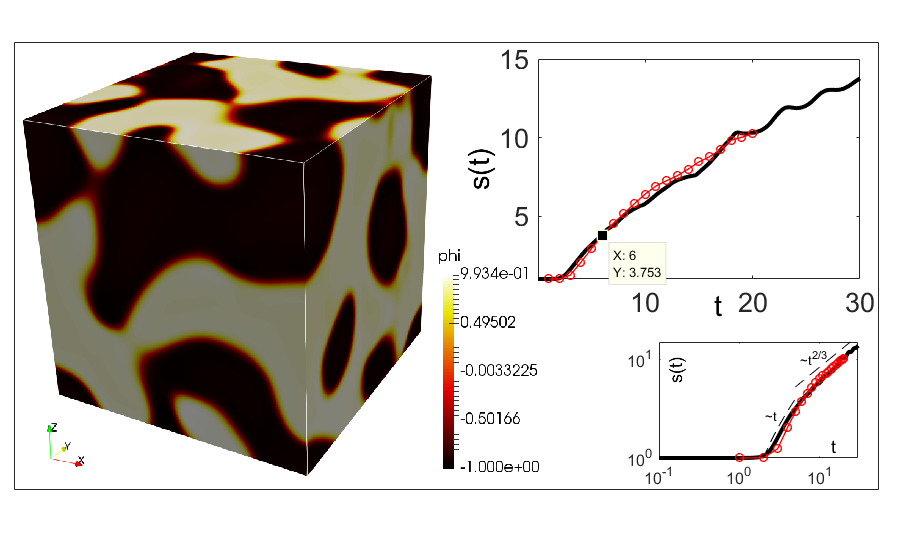}}
	\caption{Snapshot of the concentration field $\phi$ at various times for: (a) Run3D\_1 and (b) Run3D\_2. Each snapshot is accompanied by an inset showing the time evolution of $\crazyell(t)$.  Inset legend: Black solid line: two-dimensional case; red circles: three-dimensional case.   A further inset is presented in each panel showing the time evolution of $\crazyell(t)$ on a log-log scale, thereby illustrating the power-law behaviour of $\crazyell(t)$.}
\label{fig:3dsnapshots}
\end{figure}

\begin{figure}[htb!]
\centering
\includegraphics[width=0.8\textwidth]{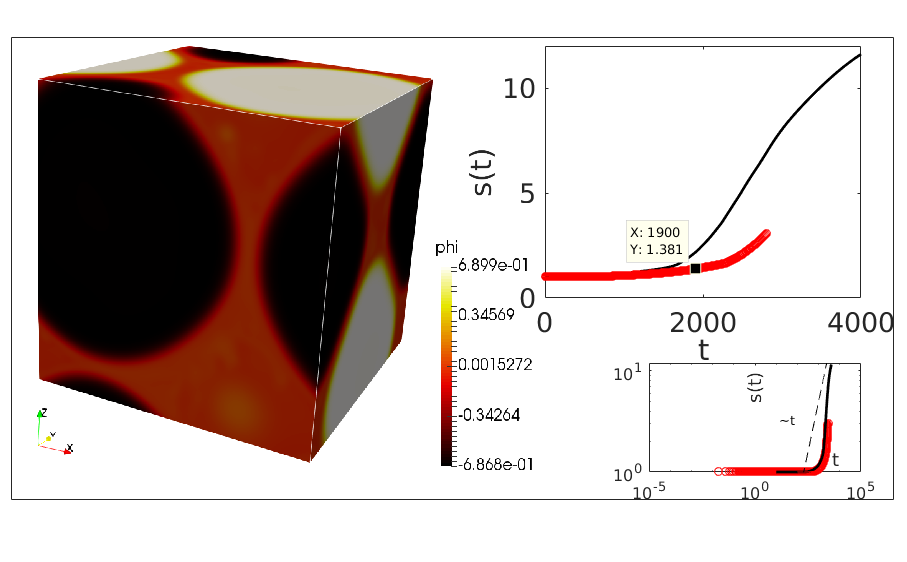}
\caption{Snapshot of the concentration field $\phi$ at various times for Run3D\_3.   Inset legend: Black solid line: two-dimensional case; thick red line/circles: three-dimensional case.  A trend line $\sim t$ in the log-log plot in the inset is included here merely to guide the eye and to illustrate the super-linear growth of $\crazyell(t)$.}
\label{fig:3dsnapshotsx}
\end{figure}

To understand whether the results presented above carry over to three dimensions, we have taken three sample test cases from the flow-pattern map in Figure~\ref{fig:flowmap} corresponding to reading off parameter cases along the diagonal, as summarized in Table~\ref{tab:3dsims}.
\begin{table}
	\centering
		\begin{tabular}{c|c|c|c}
			Case & $\nu$ & $D$ & $\Delta t$ \\
			\hline
			Run3D\_1 & $10^{-1}$ & $10^{-1}$ & $2\times 10^{-7}$\\
			Run3D\_2 & $10^{-3}$ & $10^{-3}$ & $2\times 10^{-5}$\\
			Run3D\_3 & $10^{-5}$ & $10^{-5}$ & $2\times 10^{-3}$\\
		\end{tabular}
	\caption{Summary of parameter values used for the three-dimensional simulations.  Each simulation is carried out in a periodic box of resolution $256^3$.}
	\label{tab:3dsims}
\end{table}
Sample results are shown in Figures~\ref{fig:3dsnapshots}--\ref{fig:3dsnapshotsx}.
Snapshots of the concentration $\phi$ are shown in the main part of the figures; the insets show the time evolution of $\crazyell(t)$.  The figures show that the dependence of the morphology on the parameter values is qualitatively similar in two dimensions and three dimensions.  For Run3D\_1 and Run3D\_2 the behaviour is both qualitatively and quantitatively very similar; in particular, the time evolution of $\crazyell(t)$ shows the same trend in two dimensions and three dimensions: either a crossover from viscous to inertial scaling (Run3D\_2), or the persistence of diffusive scaling for all observed times (Run3D\_1).  


In contrast, while Run3D\_3 exhibits anomalous scaling behaviour (similar to the two-dimensional analogue), the time evolution of $\crazyell(t)$ in Run3D\_3 is much slower than the analogous behaviour in two dimensions (while still not following any definitive power law).  The evolution of the morphology is correspondingly abated: the snapshot of the three-dimensional morphology at $t=1900$ in Figure~\ref{fig:3dsnapshotsx} is qualitatively very similar to the snapshot of the two-dimensional morphology at $t=900$ in the analogous two-dimensional simulation in Figure~\ref{fig:phi_anomalous2}.  The similarity between these cases is concluded on the basis that spherical domains of either $\phi\approx+1$ or $\phi\approx-1$ embedded in a matrix of well-mixed fluid, are present in both snapshots.   Thereafter, in 2D the morphology consists of bubbles of $\phi=+1$ in a sea of $\phi=-1$ -- a process referred to in Section~\ref{sec:anom} as `symmetry-breaking'.  In 3D this process is just barely visible, however finite-size effects spoil the simulation results by $t=2100$.  

\begin{figure}
	\centering
		\includegraphics[width=0.6\textwidth]{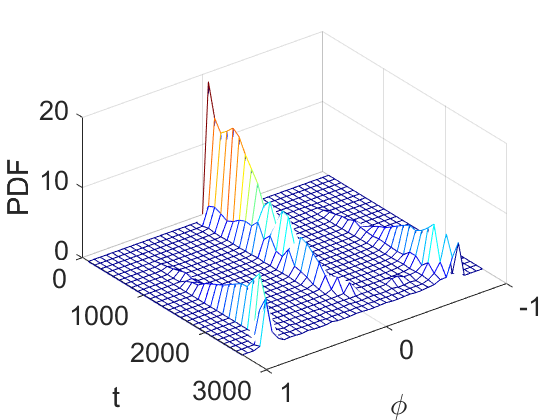}
		\caption{Time evolution of the probability distribution function of $\phi$ for the case Run3D\_3.}
	\label{fig:pdf_3D}
\end{figure}

The slowdown in the phase separation in 3D can be further seen in the time-evolution of the PDF in Figure~\ref{fig:pdf_3D}, which can be contrasted with Figure~\ref{fig:hist_compare}(b) for 2D.  In 2D, the passage to the formation of binary domains is accompanied by the `symmetry breaking' mechanism discussed earlier, which occurs at $t\approx 1500$, while in 3D the evidence for symmetry breaking (such as it is, obscured by finite-size effects) occurs at $t\approx 2000$.

\section{Conclusions}

We have performed a unified and detailed analysis of the Navier-Stokes-Cahn-Hilliard system describing phase separation in binary mixtures in the presence of flow. An extensive set of two-dimensional direct numerical simulations in which both the diffusivity and the viscosity were varied allowed us to construct a flow-pattern map outlining the relative strengths of flow and of phase separation in different parts of the parameter space. In large regions of this parameter space, the dynamics obeys the standard theory: the characteristic scale of the concentration field then grows algebraically in time, consistently with a dimensional analysis reasoning based on the balance between different terms of the Navier--Stokes-Cahn--Hilliard system. Depending on the values of diffusivity and viscosity, and on time, the system can be in the (well-known) diffusive, viscous or inertial regimes.

However, we have also shown that this standard theory does not apply in large parts of the parameter space. In particular, for low values of diffusivity and viscosity, an anomalous regime in which the coarsening is faster-than-linear in time is obtained. This regime, also characterized by the formation at intermediate times of bubbles of (almost) pure phases in a matrix of mixed fluid, had been previously reported in Stokes flows, but we also found it in flows dominated by inertia. The standard theory does not apply either in the large diffusivity-arbitrary-viscosity region of the parameter space. In this case, the diffusion is found to overwhelm unconditionally the hydrodynamics.
Some further simulations show that the same description  holds in the three-dimensional case as well, with the caveat that the passage to a fully segregated binary mixture is delayed in 3D compared to 2D.
Finally, the present work has involved the study of turbulence in the Navier--Stokes--Cahn--Hilliard system whereby the forcing is provided by the initial condition in the concentration field, which provides an excess of free energy which in turn drives the flow.  
It will be interesting to determine if the present findings carry over to the Navier-Stokes-Cahn-Hilliard system in the presence of a continuous-in-time turbulent forcing applied directly to the momentum equation.

\subsection*{Acknowledgements}

The authors acknowledge the DJEI/DES/SFI/HEA Irish Centre for High-End Computing (ICHEC), PSMN (\'Ecole Normale Sup\'erieure de Lyon), GENCI-CINES, and GENCI-IDRIS (grant x20162b6893), for the provision of computational facilities and support.  LON gratefully acknowledges a visiting professor award at the \'Ecole Centrale de Lyon and a UCD Seed Funding grant (SF 1316).

\appendix

\section{A proxy measure of the typical domain scale}
\label{sec:app:proxy}

For definiteness, the following arguments are presented for a two-dimensional system, although the result carries over to arbitrary dimensions.  Assuming that $\phi$ relaxes to $\phi=\pm 1$ rapidly in domains separated by narrow transition regions (effectively, smeared or diffuse interfaces) of typical width $\sqrt{\gamma}$, we have
\[
\int_{\Omega}\phi^2\,\mathd^2 x= \Omega_\mathrm{d}.
\]
Here $\Omega_{\mathrm{d}}$ is the total area occupied by the domains, $\Omega=[0,L]^\mydim$ is the computational domain (we also use the symbol $\Omega$ for the area of the same).   Thus,
\[
\frac{\Omega-\Omega_{\mathrm{d}}}{\Omega}=
\frac{\int_{\Omega}\mathd^2 x-\int_{\Omega}\phi^2\,\mathd^2 x}{\int_{\Omega}\mathd^2x}=
1-\langle \phi^2\rangle,
\]
where $\langle\cdot\rangle$ denotes the spatial average.  But $\Omega-\Omega_{\mathrm{d}}$ is the area $\Omega_{\mathrm{int}}$ occupied by the transition regions between the domains, with
\[
\Omega_{\mathrm{int}}=\alpha_1N_{\mathrm{int}}\sqrt{\gamma}\ell,
\]
where $\alpha_1$ is a dimensionless geometric factor (e.g. $\alpha_1=2\pi$ for circular domains of radius $\ell$), $N_{\mathrm{int}}$ is the total number of connected interfaces, and $\ell$ is the typical lengthscale of a domain.  Hence,
\[
\frac{\Omega-\Omega_{\mathrm{d}}}{\Omega}=\alpha_1\sqrt{\gamma}\ell(N_{\mathrm{int}}/\Omega)=1-\langle \phi^2\rangle.
\]
It remains to examine the density of interfaces $N_{\mathrm{int}}/\Omega$.  There is precisely one connected interface per domain, and the area of a typical domain is $\alpha_2 \ell^2$, where $\alpha_2$ is another geometric prefactor (e.g. $\alpha_2=\pi$ for circular domains of radius $\ell$), hence
\[
N_{\mathrm{int}}/\Omega=1/(\alpha_2\ell^2),
\]
hence
\[
1-\langle \phi^2\rangle=(\alpha_1/\alpha_2)\sqrt{\gamma}\ell^{-1},
\]
hence
\[
\ell\propto \crazyell(t):=\left(1-\langle \phi^2\rangle\right)^{-1}.
\]


\end{document}